\documentclass[10pt,aps,prd,showpacs,nofootinbib,twocolumn,superscriptaddress]{revtex4-2}

\usepackage{amssymb,amsmath,amsfonts}
\usepackage[usenames]{xcolor}
\usepackage[pdftex]{graphicx}
\usepackage[english]{babel}
\usepackage{amsmath,amssymb}
\usepackage{subcaption}  
\usepackage{multirow}
\usepackage{txfonts}
\usepackage{ulem}
\usepackage{ragged2e}
\usepackage[percent]{overpic}
\usepackage{mathtools}

\usepackage{lipsum}

\newcommand{\avgcs}[0]{\ensuremath{\left<c_s^2\right>~}}

\begin{document}

\title{Exceeding the conformal limit inside rotating neutron stars: \\ Implications to modified theories of gravity}

\author{Raissa F.\ P.\ Mendes}
\email{rfpmendes@id.uff.br}
\affiliation{Instituto de F\'isica, Universidade Federal Fluminense, Niter\'oi, RJ, 24210-346, Brazil.}
\affiliation{CBPF - Centro Brasileiro de Pesquisas F\'isicas, 22290-180, Rio de Janeiro, RJ, Brazil.}

\author{Caroline F. Sodré}
\affiliation{CBPF - Centro Brasileiro de Pesquisas F\'isicas, 22290-180, Rio de Janeiro, RJ, Brazil.}

\author{Felipe T. Falciano}
\affiliation{CBPF - Centro Brasileiro de Pesquisas F\'isicas, 22290-180, Rio de Janeiro, RJ, Brazil.}
\affiliation{PPGCosmo, CCE - Universidade Federal do Esp\'irito Santo, 29075-910, Vit\'oria, ES, Brazil.}

\date{\today}

\begin{abstract}
At the supranuclear densities achieved inside a neutron star, matter may exhibit extreme properties. In particular, it may be the case that a suitable average of the speed of sound squared exceeds the so-called conformal limit, i.e., $\avgcs > 1/3$, a condition that is equivalent to the positiveness of the trace of the energy-momentum tensor at the stellar center. 
This property, which holds for highly compact neutron stars obeying many (but not any) realistic equations of state, would turn these objects into interesting laboratories for tests of several scalar extensions of general relativity. 
In this paper, we investigate how rapid rotation influences the superconformality of the averaged speed of sound squared and modified gravity effects that depend thereupon, paying particular attention to scalar-tensor theories prone to the spontaneous scalarization effect.
\end{abstract}

\pacs{97.60.Jd, 04.50.Kd, 04.40.Dg} 

\maketitle

\section{Introduction}
\label{sec:intro}


At the supranuclear densities present in the core of a neutron star (NS), the behavior of nuclear matter is still poorly understood. Up to $\sim 2 n_\text{sat}$ (where $n_\text{sat} = 0.16 \, \text{fm}^{-3}$ is the number density at nuclear saturation), chiral effective field theory provides a reliable framework to compute the nuclear equation of state (EOS) \cite{Drischler:2021kxf}, while perturbative QCD methods become available at asymptotically large densities \cite{Kurkela:2009gj}. However, at the intermediate density range thought to exist in the core of NSs, different extrapolations of known physics give rise to disparate predictions for the EOS and, consequently, for NS properties \cite{Ozel:2016oaf}.

For old, low-temperature NSs, the EOS is well described by a one-parameter relation between pressure and energy density, $p = p(\epsilon)$, and the adiabatic speed of sound, $c_s^2 \equiv dp/d\epsilon$, gives a useful measure of the EOS stiffness.
For nonrelativistic nuclear matter, $c_s \ll 1$, while in the limit of asymptotically large densities, one expects that $c_s^2 \to 1/3$, which is known as the (QCD) conformal limit. At the intermediate densities present in the core of NSs, the conformal limit has been conjectured to be respected\footnote{Subsequent works demonstrated that even in the framework of holography no hard speed limit exists at finite density; see, e.g., Refs.~\cite{Hoyos2016,Ecker:2017fyh}.} \cite{Cherman:2009tw}, but observations suggest otherwise. In particular, the existence of NSs as massive as $2 M_\odot$ favors the scenario where the maximum speed of sound achieved inside a stable NS is close to the speed of light, surpassing the conformal limit \cite{Bedaque:2014sqa,Kurkela:2014vha,Alsing:2017bbc,Tews:2018kmu,Altiparmak:2022bke,Ecker:2022xxj} (e.g., $c_s^\text{max} > 0.63c$ at 99.8\% confidence level, from \cite{Alsing:2017bbc}).

A more extreme (albeit less likely) possibility is that the conformal limit is exceeded not only by the maximum speed of sound, but also on ``average'' within a NS. Let us define the density-averaged speed of sound squared as \cite{Saes:2021fzr,Saes:2024xmv}
\begin{equation} 
    \avgcs \equiv \frac{1}{\epsilon_c} \int_{0}^{\epsilon_c} c_s^2 d\epsilon = \frac{p_c}{\epsilon_c},
\end{equation} 
where we assume that $\epsilon = 0$ at the stellar surface, and the subscript $c$ denotes the central value of a given quantity. The second equality then follows immediately, as well as the relation between $\avgcs$ and the (scaled) trace anomaly defined as $\Delta = 1/3 - p/\epsilon$, evaluated at the stellar center \cite{Fujimoto:2022ohj,Marczenko:2022jhl}. Thus, asking whether the superconformality condition
\begin{equation} \label{eq:superconformal}
    \avgcs > \frac{1}{3}
\end{equation}
is verified inside (some) stable NSs is equivalent to asking whether the trace of the energy momentum becomes positive at the stellar center ($T_c = 3p_c - \epsilon_c = -3 \epsilon_c \Delta_c  >0$). Current NS data is consistent with both the scenario where $\avgcs < 1/3$ for all stable NSs and that where $\avgcs > 1/3$ for some stable NSs (see, e.g., Fig.~1 of Ref.~\cite{Fujimoto:2022ohj}). The existence of approximately universal relations between $\avgcs$ and macroscopic NS properties such as compactness, moment of inertia, and tidal deformability \cite{Saes:2021fzr,Saes:2024xmv} should assist the inference of \avgcs from future electromagnetic and gravitational-wave observations of NSs.

Interestingly, the scenario where the conformal limit is exceeded on average, as in Eq.~\eqref{eq:superconformal}, inside some NSs would not only be informing as to how QCD approaches conformality, but would also open the door to possible new tests of modified theories of gravity. For instance, in modifications to general relativity (GR) where gravity is mediated by one or more scalar fields, the most natural coupling between the scalar and matter fields is through the trace of the energy-momentum tensor, that being the simplest invariant constructed from $T^{\mu\nu}$.
The fact that the trace might change sign inside a NS may lead to new phenomenology in such theories.

One example is provided by the \textit{descreening} of NSs in chameleon-like models \cite{deAguiar:2020urb,Dima:2021pwx,deAguiar:2021bzg}. 
Modified theories of gravity designed to model the dark sector must present some natural mechanism to suppress their effects in regimes where general relativity is well tested. Such screening effects typically rely on the environmental dependence of the scalar field properties, which is a natural consequence of the coupling to the trace of the energy-momentum tensor in models including chameleon \cite{Khoury:2003aq}, symmetron \cite{Hinterbichler:2010es} and dilaton \cite{Brax:2010gi}. As the ambient density increases, but still in the nonrelativistic regime, the effects of the scalar field are suppressed; however, around the strongly interacting, relativistic nuclear matter inside NSs, the scalar field effects may be reactivated, as a consequence of the fact that the trace of the energy-momentum tensor is not a monotonically decreasing function of rest mass density.

Another example is provided by the extension of the parameter space for \textit{spontaneous scalarization} \cite{Damour:1993hw,Mendes:2014ufa,Mendes:2016fby}. In certain classes of scalar-tensor theories, a trivial equilibrium solution can always be obtained by combining a GR solution for the metric and fluid variables to a constant scalar field profile; however such GR-like solutions are not necessarily stable and may be prone to a tachyonic-like instability \cite{Harada:1997mr}. A scalarized star, with a nontrivial scalar field content, may then form as the instability develops \cite{Novak:1998rk,Mendes:2016fby}. Crucially, this tachyonic-like instability requires a negative effective mass squared, $m_\text{eff}^2 < 0$, a quantity that is proportional to the product of the trace of the energy-momentum tensor and a theory-dependent parameter ($\beta_0$), $m_\text{eff}^2 \propto - \beta_0 T$ (cf. Sec.~\ref{sec:scalarization} for details). Depending on the sign of the trace of the energy-momentum tensor, complementary parts of the theory's parameter space can therefore be accessed (i.e., $\beta_0 <0$ if $T<0$ and $\beta_0 >0$ if $T>0$).

The main purpose of the present work is to explore how rotation influences modified gravity effects that rely on condition \eqref{eq:superconformal}.
In the first part, we delve on the $\langle c_s^2 \rangle - C_e$ relation (where $C_e = M/R_e$ denotes the stellar equatorial compactness)
for rapidly rotating NSs within the framework of GR, and make an initial assessment of how rotation influences the approximate universality found in the static limit for this relation \cite{Saes:2021fzr}. We argue that the same degree of universality should hold up to moderate rotation rates, as a consequence of the fact that sequences of constant central density lie roughly along lines of constant equatorial compactness as long as the angular velocity is sufficiently below the Kepler limit. 

By analyzing the radial profile of the trace of the energy momentum tensor for rotating NSs in GR, we then attempt to build some intuition on whether rotation favors or disfavors condition \eqref{eq:superconformal} and the effects that depend thereupon. 
However, when it comes to understanding specific beyond-GR effects, rotation may act in a multifaceted manner, affecting both the fluid and the spacetime, and such intuition is not necessarily complete. Therefore, in the second part of this work, we turn to explore a specific beyond-GR effect, investigating how rotation affects the parameter space for the tachyonic instability that typically heralds spontaneous scalarization, in the $\beta_0>0$ case. Rapidly rotating, scalarized NSs were constructed in \cite{Doneva:2013qva,Doneva:2016xmf,Doneva:2018ouu,Staykov:2023ose} for uniform and differential rotation, but only in the $\beta_0 < 0$ case, for which condition \eqref{eq:superconformal} is unnecessary. In this work, we study the $\beta_0>0$ case by following a complementary approach to that employed, e.g., in Ref.~\cite{Doneva:2013qva}, namely, by performing a linear stability analysis on top of a GR background instead of explicitly constructing scalarized equilibrium solutions. We find that, for a fixed central pressure, rotation tends to amplify the parameter space where GR solutions are unstable under scalar perturbations, which is indicative of the parameter space where scalarized solutions exist. Nonetheless, since rotation tends to decrease the central pressure of the most massive NS configuration, the instability regions overall become thinner as the angular velocity increases (cf.~Fig.~\ref{fig:parameterspace}).

The remainder of the paper is organized as follows.
In Sec.~\ref{sec:cs2} we delve into the $\langle c_s^2 \rangle - C_e$ relation for rapidly rotating NSs in GR, and assess how rotation influences condition \eqref{eq:superconformal}. Section \ref{sec:implications} discusses an example of a modified gravity effect that relies on condition \eqref{eq:superconformal} and provides details on the computation of the parameter space for spontaneous scalarization when $\beta_0 >0$. Section \ref{sec:conclusions} contains our main conclusions. We use natural units, $c = G = 1$, unless indicated otherwise.


\section{$\avgcs > 1/3$ for rotating neutron stars} 
\label{sec:cs2}

\subsection{Basic definitions}

We begin by briefly stating the main definitions and strategy followed throughout this section. 
In order to compute the properties of rapidly rotating NSs in GR, we employ the $\mathsf{rns}$ code \cite{Stergioulas:1994ea}. It implements a version of the KEH scheme \cite{Komatsu:1989zz}, which is based on the iterative solution of integral representations of the relevant field equations. Employing an isotropic radial coordinate, the line element of an asymptotically flat, stationary, and axisymmetric spacetime can be cast as
\begin{equation} \label{eq:ds2}
    ds^2 = - e^{\gamma + \rho} dt^2 + e^{\gamma - \rho} \bar{r}^2 \sin^2 \theta (d\varphi - \omega dt)^2 + e^{2 \alpha} (d \bar{r}^2 + \bar{r}^2 d\theta^2),
\end{equation}
where the metric potentials $\gamma$, $\rho$, $\omega$, and $\alpha$ depend on $\bar{r}$ and $\theta$. 

Matter is described by a perfect fluid with energy-momentum tensor
\begin{equation} \label{eq:perfect_fluid}
    T^{\mu\nu} = (\epsilon + p) u^\mu u^\nu + p g^{\mu\nu},
\end{equation}
where $u^\mu$ is the fluid four-velocity, $\epsilon$ is the energy density, and $p$ is the pressure as measured by comoving observers. We extract from the numerical data the ADM mass $M$, angular momentum $J$, and baryon mass $M_b$, as well as the circumferential equatorial radius, 
\begin{equation}
    R_e = e^{\left[\gamma(\bar{r}_*,\pi/2) - \rho(\bar{r}_*,\pi/2)\right]/2} \bar{r}_* \ ,
\end{equation}
where $\bar{r}_*$ denotes the isotropic radial coordinate at which pressure vanishes (in the $\theta=\pi/2$ plane).
In our analyses, we consider rotation rates up to the mass-shedding (or Kepler) limit, i.e., until the equatorial angular velocity of the fluid elements reaches the angular velocity $\Omega_K$ of a free particle in circular orbit at the equator.

Throughout the paper, we assume that the NS is described by a cold, one-parameter equation of state, and employ three tabulated nuclear models for the EOS, 
KDE0V1 \cite{Gulminelli:2015csa,Danielewicz:2008cm,Agrawal:2005ix}, SLY9 \cite{Gulminelli:2015csa,Danielewicz:2008cm} and SKI3 \cite{Gulminelli:2015csa,Danielewicz:2008cm,Reinhard:1995zz}, with tables extracted and adapted from the CompOSE repository \cite{Typel:2013rza}.

\subsection{Universal \avgcs -- $C_e$ relation for rapidly rotating NSs}

\begin{figure}[tbh]
\centering
  \includegraphics[width=0.97\linewidth]{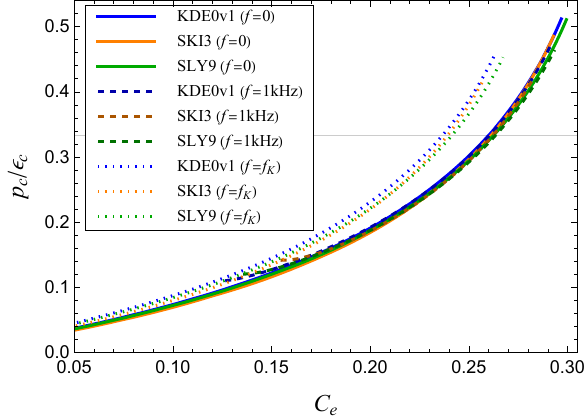}
\caption{\justifying{Ratio $p_c/\epsilon_c$ as a function of the equatorial compactness, $C_e = M/R_e$, for three EOSs and three rotation rates, $f=0$, $f=1 \text{ kHz}$, and $f = f_K$. Sequences with $f=1 \text{ kHz}$ are truncated at the mass-shedding limit. A gray line identifies configurations for which $\avgcs = p_c/\epsilon_c = 1/3$.}}
\label{fig:cs2C}
\end{figure}

In the nonrotating limit, the ratio $p_c/\epsilon_c$, or equivalently $\avgcs$, is related to the stellar compactness in an EOS insensitive way \cite{Saes:2021fzr}. This universality can be linked to a Newtonian limit that is (approximately) shared by a wide family of EOS relevant to the description of NSs, and weakens somewhat as relativistic corrections are included \cite{Saes:2024xmv}. The $\avgcs\!\! - C_e$ relation is displayed in Fig.~\ref{fig:cs2C} for the three EOSs considered in this work, for both static and rotating configurations. Two types of rotating sequences are considered: one with a fixed rotation frequency of 1 kHz -- faster than any known radio pulsar \cite{Hessels2006} -- and another at the mass-shedding rotation frequency [$f_K = \Omega_K/(2\pi)$]. 

\begin{figure}[tbh]
\centering
  \includegraphics[width=0.97\linewidth]{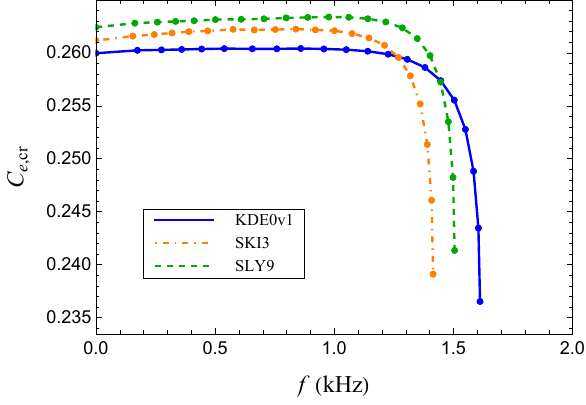}
\caption{\justifying{Equatorial compactness of the ``critical'' configuration with $\avgcs = p_c/\epsilon_c = 1/3$ as a function of the rotation frequency, for the three EOSs considered in this work. The sequences are terminated at the Kepler frequency. Up to moderately high rotation rates, the variability due to the EOS is larger than that introduced by rotation.}}
\label{fig:Ccritical}
\end{figure}

Up to moderate rotation rates, the relation between the equatorial compactness and \avgcs is very weakly dependent on rotation. 
This can be seen, in Fig.~\ref{fig:cs2C}, from the sequences spinning at 1 kHz, which only deviate significantly from the static cases when this frequency approaches the corresponding Kepler limit, at values of $C_e \lesssim 0.13-0.16$, depending on the EOS. From Fig.~\ref{fig:cs2C} one sees that, as the mass-shedding limit is approached, the equatorial compactness decreases along a constant $\avgcs$ sequence. This behavior is better visualized in Fig.~\ref{fig:Ccritical}, where the equatorial compactness of the ``critical'' configuration with $\avgcs = 1/3$ is shown as a function of the rotation frequency, up to the Kepler limit. For rotation frequencies $f \lesssim 1 \text{ kHz}$, the critical compactness for which $\avgcs = 1/3$ is not significantly altered, with the variability due to the EOS being larger than that introduced by such moderate rotation rates.

\begin{figure*}[thb]
\centering
\includegraphics[width=\linewidth]{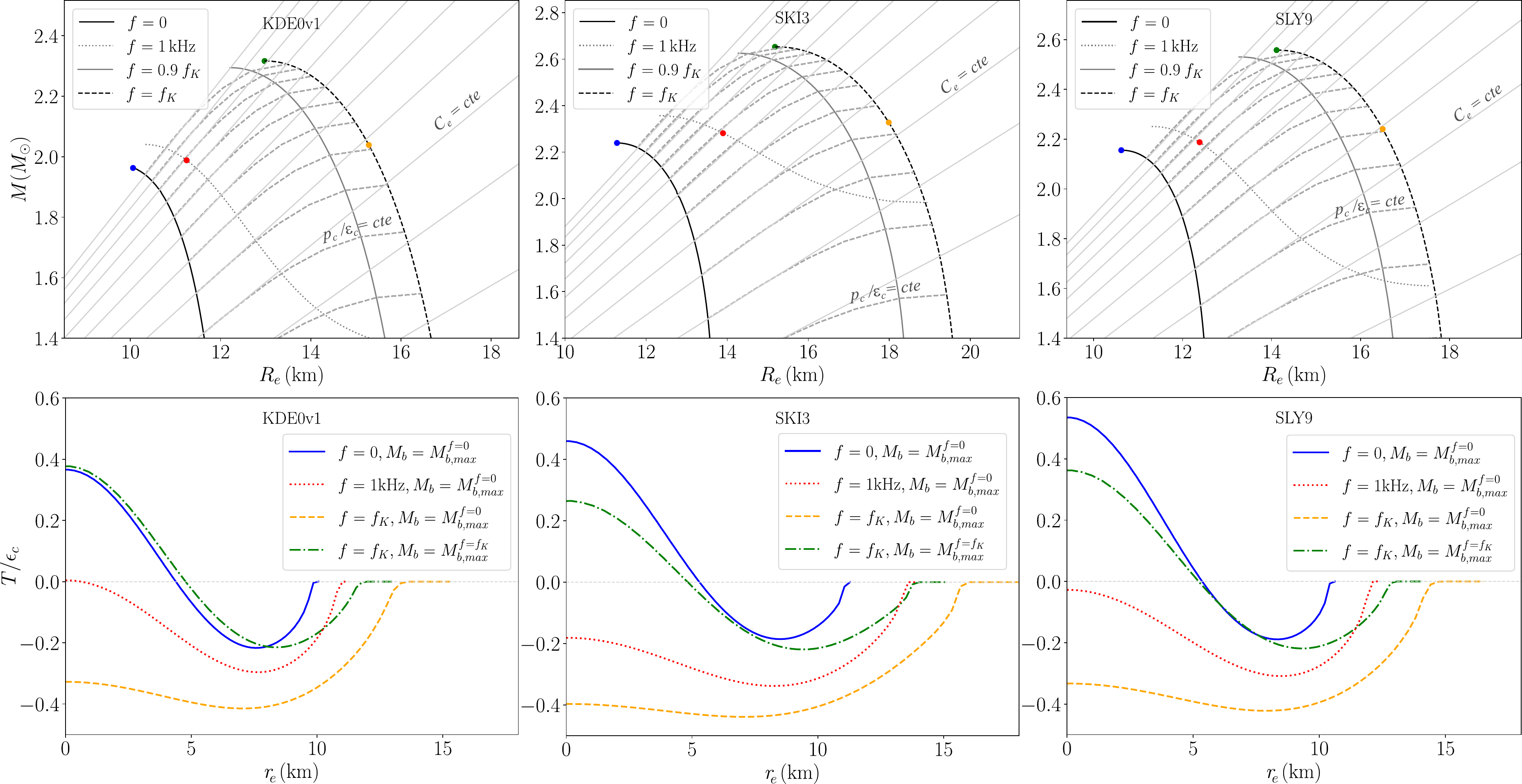}
\caption{\justifying
\textit{Upper panels}: Sequences of equilibrium configurations following the KDE0v1 (left), SKI3 (middle), and (right panel) SLY9 EOS, with rotation frequencies $f=0$, $f=1 \text{ kHz}$, $f= 0.9 f_K$ and $f=f_K$, as well as sequences with constant values of $\avgcs = p_c/\epsilon_c$ (gray dashed curves). Up to moderate rotation rates, the later lie along lines with constant equatorial compactness. 
\textit{Bottom panels}: Profile of the trace of the energy-momentum tensor, $T = 3p-\epsilon$, in the equatorial plane, as a function of a circumferential radial coordinate ($r_e = e^{\left[\gamma(\bar{r},\pi/2) - \rho(\bar{r},\pi/2)\right]/2} \bar{r}$), for four configurations obeying the KDE0v1 (left), SKI3 (middle), and SLY9 (right) EOS: (i) the maximum-mass, nonrotating solution, which has a baryonic mass of $M_{b,\text{max}}^{f = 0} \approx 2.55 M_\odot$, (ii) a solution with the same rest mass, but with a rotation frequency of 1kHz, (iii) a solution with the same rest mass, but rotating at the Kepler frequency, and (iv) the maximum mass solution rotating at the Kepler frequency. These four configurations are highlighted as colored dots in the upper panels. The trace of the energy-momentum tensor is rescaled by the central energy density, so that, at $r_e=0$, it reduces to $3\avgcs - 1$.}
\label{fig:MRe}
\end{figure*}

The results described above should be expected, as it is known that, up to moderate rotation rates, the equatorial compactness is roughly constant along a constant central-density sequence \cite{Konstantinou:2022vkr}. If one assumes a barotropic EOS, which fixes pressure as a function of energy density, then \avgcs is a function of $\epsilon_c$ alone, and constant \avgcs amounts to constant $\epsilon_c$. The upper panels of Fig.~\ref{fig:MRe} display $M - R_e$ sequences for the three EOSs we consider. One can clearly see that sequences of constant $\avgcs$ track the lines of constant equatorial compactness up to moderately high rotation rates.

\subsection{Trace profile for rotating configurations}

As discussed in Sec.~\ref{sec:intro}, a main focus of this work is on modified gravity effects that rely on condition \eqref{eq:superconformal}, or, in other words, on the existence of a region inside NSs where $T = 3p - \epsilon > 0$, and in assessing how they are affected by rotation. Before that, let us first attempt to build some intuition on how rotation may influence effects that depend on $T>0$.

The radial profile of the trace of the energy-momentum tensor in the equatorial plane is shown in the lower panels of Fig.~\ref{fig:MRe} for three EOSs and four configurations: (i) the maximum-mass static solution, (ii) a rotating solution with the same baryon mass as (i) and a 1 kHz rotation frequency, (iii) a solution with the same baryon mass as (i) but rotating at the mass-shedding limit, and (iv) the maximum-mass solution rotating at the mass-shedding limit\footnote{Although in the presence of rotation the axisymmetric instability does not set in at the maximum-mass configuration, this solution is typically close to the marginally stable one, and will be considered as equivalent to that in practice \cite{Stergioulas:1994ea}.}. These four configurations are highlighted as dots in the mass-radius relations in the upper panels of Fig.~\ref{fig:MRe}. If one fixes the baryon mass and increases the rotation rate, the equatorial radius increases and the equatorial compactness decreases, as well as the ratio $p_c/\epsilon_c$. From the lower panels of Fig.~\ref{fig:MRe} one sees that the effect of spinning up the maximum-mass static solution while keeping fixed its baryon mass is to decrease the region where $T>0$, up to the point where it ceases to exist. Still, for super-massive configurations, i.e., configurations with a larger rest mass than that supported in the static limit, a sizable region with $T>0$ may still be present. 

A general expectation that can be drawn from this analysis is that rotation would tend to decrease the $T>0$ region inside a NS if it was spun (e.g. by accretion) but without a significant increase in baryon mass. On the other hand, an increase in mass large enough to keep the equatorial compactness constant with respect to the static case should keep the $T>0$ region roughly unchanged compared to that case, as long as moderate rotation rates are considered. Naturally, rotation may affect modified gravity effects in multiple ways, acting upon both the fluid and the spacetime, and the intuition build here is not necessarily complete. In the next section, we turn to consider a specific beyond GR effect and how it is affected by rotation.

\section{Implications for modified theories of gravity}
\label{sec:implications}

\subsection{Framework}
\label{sec:stt_framework}

The existence of a coupling between the scalar degree of freedom and the trace of the energy-momentum tensor is generally expected in scalar extensions of GR -- or scalar-tensor theories (STTs) --, as $T = g_{\mu\nu} T^{\mu\nu}$ is the simplest scalar quantity that can be built from $T^{\mu\nu}$. As a consequence, NS properties may be qualitatively different if $T$ changes sign in the stellar interior, or, equivalently, if $\avgcs > 1/3$. 

In what follows, we shall consider a restricted set of scalar-tensor theories (STTs), which is nonetheless sufficiently diverse in the phenomenology predicted for NSs. Specifically, we consider STTs described by the (``Einstein frame'') action
\begin{align} \label{eq:action}
    S[g_{\mu\nu};\phi;\Psi_m] &= \frac{1}{16\pi} \int{d^4 x \sqrt{-g} \left[ R - 2 g^{\mu\nu} \nabla_\mu \phi \nabla_\nu \phi - V(\phi) \right]} \nonumber \\
    &+ S_m[\Psi_m;A(\phi)^2 g_{\mu\nu}],
\end{align}
where $g\coloneqq \det(g_{\mu\nu})$, $R$ is the Ricci scalar, and $\Psi_m$ denote matter fields, governed by the action $S_m$, and with energy-momentum tensor
\begin{equation}
    T_{\mu\nu} = - \frac{2}{\sqrt{-g}} \frac{\delta S_m}{\delta g^{\mu\nu}}.
\end{equation}
A theory within this class is specified by the choice of the two functions of the scalar field, the potential $V(\phi)$ and the conformal coupling $A(\phi)$, which determines the effective (``Jordan frame'') metric $\tilde{g}_{\mu\nu} \coloneqq A(\phi)^2 g_{\mu\nu}$ to which matter fields are universally coupled. A Jordan-frame energy-momentum tensor can also be naturally defined as $\tilde{T}_{\mu\nu} \coloneqq -2 (-\tilde{g})^{-1/2} \delta S_m /\delta \tilde{g}^{\mu\nu} = A(\phi)^{-2} T_{\mu\nu}$, and this quantity can be shown to be covariantly conserved: $\tilde{\nabla}^\nu \tilde{T}_{\mu\nu} = 0$.

The field equations that stem from the action \eqref{eq:action} are 
\begin{equation} \label{eq:sst_fe1}
    G_{\mu\nu} = 8\pi T_{\mu\nu} + 2 \nabla_\mu \phi \nabla_\nu \phi - g_{\mu\nu} \nabla_\rho \phi \nabla^\rho \phi - \frac{1}{2} g_{\mu\nu} V(\phi) 
\end{equation}
and
\begin{equation}\label{eq:sst_fe2}
    \nabla^\mu \nabla_\mu \phi = \frac{1}{4} \frac{dV}{d\phi} - 4\pi T \frac{d \ln A}{d\phi},
\end{equation}
where one sees that the scalar field dynamics depends both on the theory and on the matter content, through the trace $T$ of the energy-momentum tensor. 
As we deal with NSs, we consider the energy-momentum tensor to have a perfect fluid form \eqref{eq:perfect_fluid}, and specify the equation of state in the Jordan frame, i.e., $\tilde{p} = \tilde{p}(\tilde{\rho})$, for in this frame the first law of thermodynamics holds in its usual form. 

In what follows, we investigate the case of a non-self-interacting scalar field, and therefore set $V(\phi) = 0$. Moreover, we assume that far away from the NS, the field settles to some cosmological value $\phi_0$ such that $dA/d\phi|_{\phi_0} = 0$. Mathematically, this condition implies that an exact solution to the scalar-tensor field equations (\ref{eq:sst_fe1}) and (\ref{eq:sst_fe2}) can always be obtained by combining a constant scalar field profile, $\phi = \phi_0$, with metric and fluid configurations that obey the GR field equations -- we shall call this a ``GR-like solution''. Physically, this condition is acceptable since it renders the STT indistinguishable from GR in the weak field limit, therefore evading solar system constraints \cite{Damour:1995kt}.

\subsection{The parameter space for spontaneous scalarization in the presence of rotation} 
\label{sec:scalarization}

Although a GR-like solution for NSs with a constant scalar field profile always exists in non-self-interacting STTs with $dA/d\phi|_{\phi_0} = 0$, these solutions are not necessarily stable under scalar field perturbations \cite{Harada:1997mr,Mendes:2014ufa}. Often, instability of GR-like solutions sets in at the same critical point where additional equilibrium solutions, with a nontrivial scalar field profile, branch off the GR sequence\footnote{This can be understood since, at the critical point, scalar field perturbations around a GR-like solution are marginally stable, and thus possess a zero-frequency scalar mode. Since their time dependence vanishes, these solutions can be reinterpreted as stationary (equilibrium) solutions of the full system \eqref{eq:sst_fe1}-\eqref{eq:sst_fe2} but with vanishingly small scalar field amplitude. This is precisely what characterizes the boundary of the parameter space for spontaneous scalarization (see also \cite{Pani:2010vc} for the argument in the static case).} \cite{Harada:1998ge}. 
When these scalarized solutions are stable, they describe the physical state of the star beyond the critical point; see \cite{Doneva:2022ewd} for a review of this ``spontaneous scalarization'' effect. 

As argued above, one typically expects the boundary of the existence region for scalarized equilibrium solutions to coincide with that for the linear instability of GR-like configurations under scalar field perturbations. Thus, in this section we investigate the latter as a proxy to the former\footnote{The general argument for the coincidence of the \textit{boundaries} of these regions does not necessarily imply that the full regions coincide (see, e.g., Fig.~9 of Ref.~\cite{Mendes:2016fby} for an example otherwise), although that is usually the case for stable scalarized solutions.}. 

If we set $\phi = \phi_0 + \delta \phi$, and consider $\delta \phi$ to be sufficiently small, so that $O(\delta \phi^2)$, backreaction effects can be neglected, then the scalar field perturbation evolves on a GR background according to
\begin{equation}\label{eq:scalar_pert}
    \nabla^\mu \nabla_\mu \delta \phi = - 4\pi \beta_0  T \delta \phi,
\end{equation}
where covariant derivatives refer to the GR background metric and we defined the parameter 
\begin{equation} \label{eq:beta0}
\beta_0 \coloneqq \left. \frac{1}{A(\phi_0)} \frac{d^2 A}{d\phi^2}\right|_{\phi_0}.
\end{equation}
The evolution of linear scalar field perturbations depends on the theory in case solely through the parameter $\beta_0$, and on the matter content through the trace of the energy-momentum tensor. When the product $\beta_0 T$ is positive, Eq.~(\ref{eq:scalar_pert}) may allow for unstable solutions, as a negative effective mass squared ($m_\text{eff}^2 = - 4 \pi \beta_0 T$) can be associated with possible tachyonic-like instabilities. In what follows, we restrict attention to the case where $\beta_0 > 0$, for which the existence of unstable scalar field perturbations depends on the existence of a region inside the star where $T>0$. The $\beta_0 < 0$ case has been previously investigated in Ref.~\cite{Doneva:2013qva} (see also \cite{Doneva:2016xmf,Doneva:2018ouu,Staykov:2023ose} for generalizations) through the explicit construction of rotating, scalarized equilibrium solutions. 

We numerically solve Eq.~(\ref{eq:scalar_pert}) in the background spacetime of a rapidly rotating NS, described by the line element (\ref{eq:ds2}). Given the axial symmetry of the background spacetime, perturbations can be decomposed as
\begin{equation} 
    \delta \phi (t,\bar{r},\theta,\varphi) = \sum_m \Psi_m (t,\bar{r}, \theta) e^{im\varphi}.
\end{equation}
In what follows we concentrate on the axisymmetric mode $\Psi_0$, which obeys the following $1+2$ linear partial differential equation:
\begin{align} \label{eq:PDEPsi0}
    - e^{-\gamma -\rho} \partial_t^2 \Psi_0 + e^{-2\alpha} \partial_{\bar{r}}^2 \Psi_0 + e^{-2\alpha} \left( \frac{2}{\bar{r}} + \partial_{\bar{r}} \gamma \right) \partial_{\bar{r}} \Psi_0 + \nonumber \\ \frac{e^{-2\alpha}}{r^2} \left[ \partial_\theta^2 \Psi_0 + (\cot\theta + \partial_\theta \gamma) \partial_\theta \Psi_0 \right] = - 4 \pi \beta_0 T \Psi_0.
\end{align}
The background spacetime and fluid configurations are computed through the $\mathsf{rns}$ code \cite{Stergioulas:1994ea}, and Eq.~(\ref{eq:PDEPsi0}) is numerically integrated through the method of lines, employing a centered second-order finite-difference approximation for radial and angular derivatives, and a Runge-Kutta 3 scheme for the time integration. We use a regular grid for the radial and angular variables (the latter of which we take as $\mu = \cos\theta$), with a time step determined from the Courant–Friedrichs–Lewy condition. 
We assume a time-symmetric initial condition, such that $\partial_t \Psi_0|_{t=0} = 0$ and 
\begin{equation} \label{eq:initial_condition}
\Psi_0|_{t=0} = A e^{-(\bar{r} - \bar{r}_0)^2/\sigma^2}
\end{equation}
is a Gaussian shell with amplitude $A$ and width $\sigma \sim \bar{r}_*$ centered at $\bar{r}_0 \sim  3 \bar{r}_*$, where we recall that $\bar{r}_*$ denotes the isotropic radial coordinate at which pressure vanishes in the equatorial plane. 

The following boundary conditions are imposed. At $\bar{r} = 0$, we set $\partial_{\bar{r}} \Psi_0 | _{\bar{r}=0}= 0$, as required for a regular solution, and implement this condition using a one-sided second-order finite-difference approximation for the radial derivative. As $\bar{r} \to \infty$, outgoing boundary conditions are required, i.e.,  $\lim_{\bar{r}\to\infty} \Psi_0 = \Psi_0 (t - \bar{r})$. However, since the domain is finite, this condition is enforced (in the form $\partial_t \Psi_0 |_{\bar{r}_\text{out}} = - \partial_{\bar{r}} \Psi_0|_{\bar{r}_\text{out}}$) at a finite radial coordinate, typically at $\bar{r} = \bar{r}_\text{out} \sim 20 \bar{r}_*$. Given that the outer boundary of the radial domain is relatively close to the star, some reflection coming from the boundary is expected. Still, since our main focus will be on the identification of unstable solutions, such small reflections do not affect our results appreciably. At the boundaries of the angular domain, $\theta = 0$ and $\theta = \pi$, we impose that $\partial_\theta \Psi_0|_{\theta = 0,\pi} = 0$. This condition comes from the fact that the $m=0$ mode is symmetric under the transformation $\varphi \to -\varphi$, so it is only possible for derivatives to be continuous across the poles if they vanish there \cite{Barack:2007jh}. 

\begin{figure}[th]
\centering
  \includegraphics[width=0.97\linewidth]{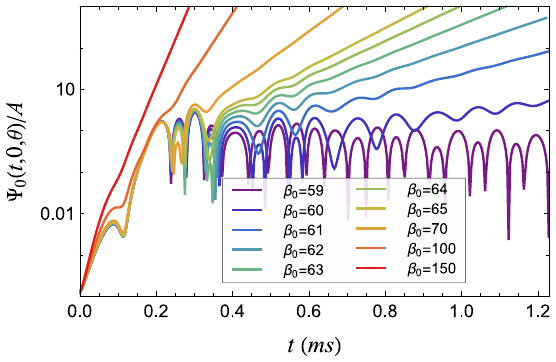}
\caption{\justifying{Time dependence of the axisymmetric mode $\Psi_0$ extracted at the center of the star. The background spacetime corresponds to a rapidly rotating NS in GR, described by the SLY9 EOS, with a central pressure $p_c/c^2 = 9.483 \times 10^{14} \text{g}/\text{cm}^3$, and rotating at the Kepler frequency ($\Omega = \Omega_K$). The perturbation is normalized by the amplitude $A$ of the initial pulse (\ref{eq:initial_condition}). Several values of $\beta_0 > 0$ are considered.}}
\label{fig:sample}
\end{figure}

\begin{figure}[th]
\centering
  \includegraphics[width=0.97\linewidth]{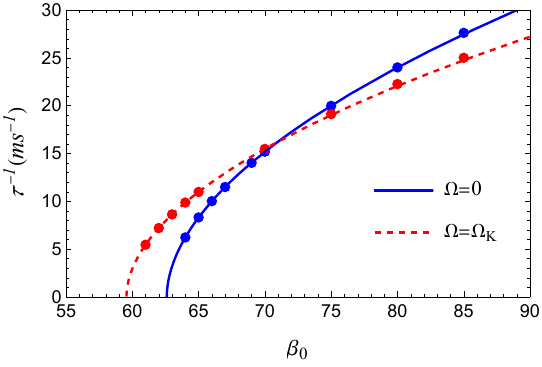}
\caption{\justifying{Inverse instability timescale as a function of $\beta_0$. The background spacetime corresponds to a NS described by the SLY9 EOS, with the same EOS and central pressure as in Fig.~\ref{fig:sample}, and an angular velocity $\Omega = 0$ (solid blue), and $\Omega = 0.9 \Omega_K$ (dashed red). Dots correspond to values extracted from simulations, as explained in the text, while curves represent best fits of the form (\ref{eq:fit}). The critical value of $\beta_0$ for which $\tau^{-1} = 0$ is extrapolated to be $\beta_\textrm{cr} \approx 62.6$ in the $\Omega=0$ case and $\beta_\textrm{cr} \approx 59.6$ for $\Omega = \Omega_K$.}}
\label{fig:timescale}
\end{figure}

A sample of the simulations we performed is presented in Fig.~\ref{fig:sample}, where we show the time evolution of $\Psi_0$, evaluated at the stellar center, for a fixed rotating NS background and several values of $\beta_0>0$. Given a central pressure that is high enough so that $T_c = 3p_c - \epsilon_c > 0$, one sees that, as $\beta_0$ increases, scalar field perturbations eventually become unstable. For those cases, the perturbation grows exponentially at late times, with a time dependence $\sim \exp(t/\tau)$, and the timescale $\tau$ of the instability decreases as $\beta_0$ increases. We aim to identify the critical value of $\beta_0$ (which we denote as $\beta_\text{cr}$) for which a given GR-like solution is marginally unstable under scalar field perturbations. For that purpose, we extract, from the late-time data for each unstable simulation, the corresponding instability timescale $\tau$. The inverse instability timescale is depicted in Fig.~\ref{fig:timescale} as a function of $\beta_0$ for the (unstable) simulations displayed in Fig.~\ref{fig:sample}.
We fit the data thus obtained with an ansatz of the type 
\begin{equation} \label{eq:fit}
\tau^{-1} \propto (\beta_0 - \beta_\text{cr})^{n},
\end{equation}
and the parameters $\beta_\text{cr}$ and $n < 1$ of the best fit are extracted. 

\begin{figure}[th]
\centering
  \includegraphics[width=0.95\linewidth]{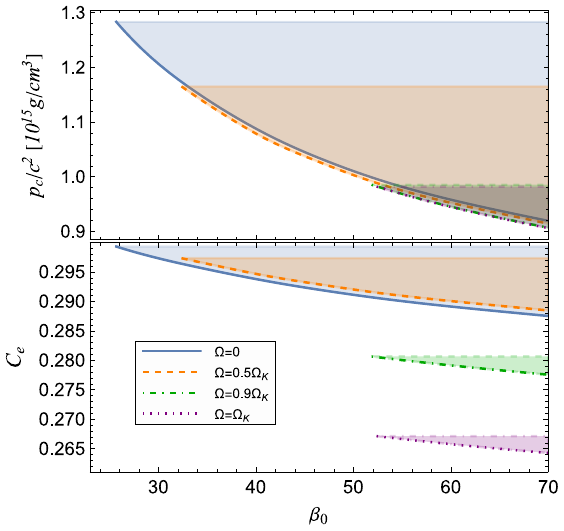}
\caption{\justifying{Parameter space where GR-like rotating NSs, described by the SLY9 EOS, are unstable under scalar field perturbations in STTs characterized by the parameter $\beta_0 >0$. In the upper panel the NS is characterized by its central pressure, and in this case all curves share the same asymptote ($\beta_0 \to \infty$ for $p_c/c^2 \approx 0.53 \times 10^{15} \text{g/cm}^3$ for this EOS). In the bottom panel, the NS is characterized by its equatorial compactness. In both cases, curves are cut at the values of $p_c$ or $C_e$ corresponding to the maximum mass configuration. Rotating configurations with $\Omega = 0.5 \Omega_K$, $\Omega = 0.9 \Omega_K$, and $\Omega = \Omega_K$ are considered, as well as the static case.}}
\label{fig:parameterspace}
\end{figure}

This procedure is repeated for several values of the central pressure, and the result is shown in Fig.~\ref{fig:parameterspace}. The shaded regions in the $\beta_0$ -- $p_c$ or $\beta_0$ -- $C_e$ diagrams correspond to (GR-like) NS configurations that would be unstable under scalar field perturbations in STTs with those values of $\beta_0$. These regions are indicative of the parameter space for spontaneous scalarization, as previously discussed. All curves are cut at the values of $p_c$ (or $C_e$) corresponding to the maximum mass configuration, which approximately marks the onset of the axisymmetric instability in GR. Since rotation tends to decrease the central pressure and the compactness of the most massive NS (cf. Fig.~\ref{fig:MRe}), the instability regions in Fig.~\ref{fig:parameterspace} naturally become thinner as the angular velocity $\Omega$ increases. 

If the central pressure is fixed, our results show that rotation tends to slightly amplify the range of values for $\beta_0$ leading to unstable configurations, as can be also seen in Fig.~\ref{fig:timescale}. Still, as visible from the same plot, for high enough values of $\beta_0$ and fixed $p_c$, rotation increases the instability timescale, thus having somewhat mixed effects. 

On the other hand, if one fixes the equatorial compactness and consider moderate rotation rates, one sees that rotation slightly decreases the range of values for $\beta_0$ leading to unstable configurations, a result in agreement with that obtained in Ref.~\cite{Mendes:2014vna} for slowly rotating thin shells (cf. Fig.~4 of that reference). As the angular velocity approaches the Kepler limit, the compactness substantially decreases along a constant central-pressure sequence (cf.~upper panels in Fig.~\ref{fig:MRe}), leading to the separation of the instability regions seen in Fig.~\ref{fig:parameterspace}. 

Although our main focus in this work regards STTs with $\beta_0 > 0$, for which spontaneous scalarization relies on the fulfillment of condition \eqref{eq:superconformal}, we show results for the $\beta_0 < 0$ case in the Appendix, which are compared with those of Refs.~\cite{Doneva:2013qva}.

\section{Conclusions} \label{sec:conclusions}

The way the speed of sound approaches the QCD conformal limit ($c_s^2 = 1/3$), and its behavior at the intermediate densities present in the interiors of NSs, not only are highly informative of the properties of nuclear matter in this regime, but may promote some of these objects to interesting probes of scalar-field extensions of GR.
This is particularly true if the nuclear EOS allows for NSs where the speed of sound, when averaged across the energy-density range present in the NS interior, exceeds the conformal limit, $\avgcs > 1/3$. In this work, we have investigated how this (averaged) superconformality condition, and modified gravity effects that depend on it, are affected by rotation. 

We began, in the framework of GR, by investigating how the approximate universality previously found between $\avgcs$ and the stellar compactness \cite{Saes:2021fzr,Saes:2024xmv}, is affected by rotation. We found that the $\avgcs-C_e$ relation, where $C_e$ denotes the equatorial compactness, is preserved up to moderate rotation rates (cf.~Fig.~\ref{fig:cs2C}). This is related to the fact that the equatorial compactness is roughly constant along a constant central-density sequence, as long as one is sufficiently below the mass-shedding limit (cf.~the upper panels in Fig.~\ref{fig:MRe}). We note that the pulsars currently considered for NICER pulse profile analyses have spin frequencies $< 300$ Hz \cite{Bogdanov:2019qjb}, while future experiments such as STROBE-X \cite{strobex} and eXTP \cite{expt} will target millisecond pulsars. Attempts to statistically infer $\avgcs$ from $C_e$ measurements from these missions could safely use the static $\avgcs$ -- $C$ relation for pulsars with typical compactness ($C \gtrsim 0.2$).

Next, still in the framework of GR, we attempted to gain some intuition of how rotation may affect beyond-GR effects that depend on condition \eqref{eq:superconformal} by analyzing the radial profile of the trace of the energy-momentum tensor (the central value of which is proportional to $\avgcs - 1/3$) inside rotating NS configurations (cf.~the bottom panels in Fig.~\ref{fig:MRe}). A general expectation drawn from this analysis is that spinning up a NS without a significant increase in its mass should lead to the quenching of the region inside the star where $T>0$, and that rapidly spinning NSs with $\avgcs>1/3$ are most likely supermassive (i.e., more massive than the most massive, static configuration). In our work, this initial assessment was made using three realistic EOS, but a more thorough analysis could be done by covering more of the EOS uncertainty band.

We proceeded to contrast our general expectations with the case of a particular scalar-tensor theory, described by the action \eqref{eq:action} with $V(\phi) = 0$ and $A(\phi)$ such that $dA/d\phi|_{\phi_0} = 0$, where $\phi_0$ is the present-time cosmological value of the scalar field. In this case, a GR solution for the metric and fluid quantities can always be promoted to a solution in the full theory as long as $\phi = \phi_0$; however, this GR-like solution may not be linearly stable under scalar field perturbations. The boundary of the region in parameter space for this linear instability is understood to coincide with that for the spontaneous scalarization effect. Since we are dealing with linear perturbations, not the entire content of the conformal coupling $A(\phi)$ is relevant, and the behavior of these perturbations is completely governed by a single theory parameter $\beta_0$, defined in Eq.~\eqref{eq:beta0}. A necessary condition for the linear instability of GR-like configurations is then the positiveness of the product $\beta_0 T$; when $T>0$ (as $\avgcs>1/3$), linear instability can occur in the $\beta_0 >0$ region of the theory's parameter space. In Sec.~\ref{sec:scalarization} we investigated how this parameter space is affected by rotation; see Fig.~\ref{fig:parameterspace} for the main result of this section. 
Our results conform to the previously built intuition, as far as, for instance, low rotation rates do not sensibly affect the parameter space, while for high rotation rates, a thinner range of stellar configurations are prone to this linear instability. Other observations, such as the fact that, for a fixed central pressure, rotation (slightly) amplifies the range of values of $\beta_0$ leading to instability, while decreasing the instability timescale for sufficiently large values of $\beta_0$ (cf. Fig.~\ref{fig:timescale}) are thought to be case sensitive. Naturally, rotation affects the scalar field evolution in multiple ways, not only through the trace profile of the rotating fluid configurations. We highlight that the results in this section were obtained purely for the axisymmetric mode, with $m=0$, but the nonaxisymmetric case could be tackled in an analogous manner.

\acknowledgments
This work was partially supported by the National Council for Scientific and Technological Development (CNPq), and by the Carlos Chagas Filho Research Support Foundation (FAPERJ).


\appendix

\section{The parameter space for spontaneous scalarization in the presence of rotation: $\beta_0 < 0$ case}

\begin{figure}[b]
\centering
  \includegraphics[width=0.95\linewidth]{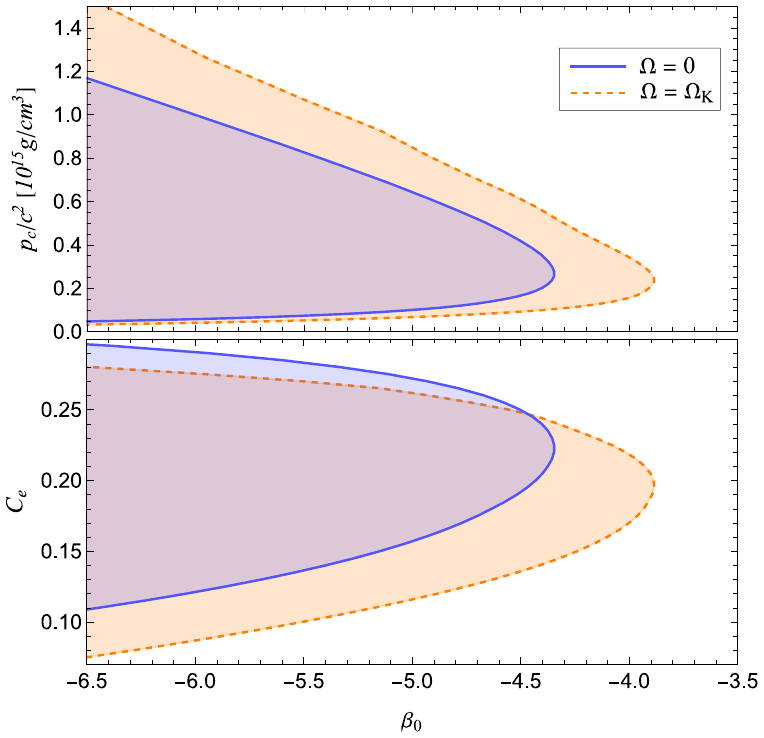}
\caption{\justifying{Parameter space where GR-like NSs, described by the SLY9 EOS, are unstable under scalar field perturbations in STTs characterized by the parameter $\beta_0 <0$. The NS is identified by its central pressure in the upper panel and by its equatorial compactness in the bottom panel. We consider configurations rotating at the Kepler limit (dashed orange), as well as the static case (solid blue) for comparison.}}
\label{fig:parameterspaceminus}
\end{figure}

For completeness, we display in Fig.~\ref{fig:parameterspaceminus} the parameter space for spontaneous scalarization for the SLY9 EOS (same employed in Fig.~\ref{fig:parameterspace}), but for $\beta_0 < 0$. The exact same procedure described in the main text was adopted in this case. One sees that, for a fixed central pressure, rotation tends to amplify the range of values of $\beta_0 < 0$ leading to unstable scalar modes of GR-like configurations. The largest value of $\beta_0 <0$ possibly leading to the existence of such unstable scalar modes increases from $\approx -4.34$ in the static case to $\approx -3.88$ at the mass-shedding limit. Our results are in agreement with those reported in Ref.~\cite{Doneva:2013qva}, where the parameter space of spontaneous scalarization was accessed by the explicit construction of equilibrium solutions; there they find that scalarized solutions can exist for $\beta_0 \lesssim -3.9$. We emphasize that this stretching of the parameter space for spontaneous scalarization is only relevant for angular velocities close to the mass-shedding limit; for the moderate rotation rates that characterize, e.g., all known pulsars, the effects due to rotation are much more discrete. 
Finally, note that the increase, due to rotation, in the range of $\beta_0$ leading to scalarization for a fixed $p_c$ is of the same order both for $\beta_0 <0$ and $\beta_0 >0$, the scales of Figs.~\ref{fig:parameterspace} and \ref{fig:parameterspaceminus} being radically different.


\bibliography{lib}

\end{document}